\documentclass[oneside,letterpaper]{amsart}

\usepackage{geometry}
\usepackage{graphicx}
\usepackage{url}

\begin{document}

\title{A Dual Non-abelian Yang-Mills Amplitude in Four Dimensions}
\author{J. Wade Cherrington$^1$}
\author{J. Daniel Christensen$^2$}
\maketitle

\vspace*{-10pt}
\begin{center}
\small
$^1$Department of Applied Mathematics, University of Western 
Ontario, London, Ontario, Canada \newline
$^2$Department of Mathematics, University of Western 
Ontario, London, Ontario, Canada
\end{center}

\begin{abstract}
We derive an explicit formula for the vertex amplitude of dual $SU(2)$ Yang-Mills theory in 
four dimensions on the lattice, and provide an efficient algorithm  (of order $j^4$)  for its computation.
This opens the way for both numerical and analytical development of dual methods, previously limited to 
the case of three dimensions.  
\end{abstract}

\section{Introduction}
Since their inception in the early 1970's, quantum field theories formulated on the lattice have 
been mapped to dual theories --- models mathematically equivalent to the original formulation,
but with different degrees of freedom. In many cases, configurations in the dual model 
can be interpreted as extended geometric structures such as closed sheets or self-avoiding polymers living 
on the space-time lattice~\cite{fermions,Fort,KST}; such structures can in turn be interpreted as 
histories of the physical degrees of freedom of the theory~\cite{ArocaPathint}.
Recently, the concept of dual methods as an alternative framework for computations in lattice gauge 
theory has seen a renewal of interest~\cite{ChandraDual,WolffWorm}. 
Apart from the geometric insight provided by the dual theories and their use of gauge-invariant 
physical degrees of freedom, dual methods may provide a source of novel algorithms
with attractive numerical performance in some regimes.  From early work (e.g.~\cite{Savit80}) it was found that 
abelian gauge theories are particularly tractable in their dual form. 
The abelian case was initially a source of many dual results and continues to provide a practical 
alternative to conventional numerical simulations~\cite{Jersak,Panero2004,Panero2005,PollyWiese,Zack98PhysRev}. 
With the above motivations in mind, we turn to the dual model of $SU(2)$ Yang-Mills theory on the 
four-dimensional lattice. This case is an important proving ground for the viability of dual non-abelian
method, and one that has long been out of reach for practical calculations.

In analogy with dual abelian constructions, a dual formulation for pure non-abelian Yang-Mills 
lattice gauge theory (LGT) in four dimensions can be constructed from the character expansion 
of a local lattice action  and has been known for decades (e.g.\ see~\cite{MandM} and references 
therein). However, the na\"{\i}ve application of character expansion leads to a non-local 
expression for the amplitude --- a group integral over the entire lattice. It has since been 
realized by several authors that this group integral, which involves products of characters, can 
be expanded into intertwiner labellings at lattice edges, such that the dual amplitude becomes a product
of amplitudes \emph{local} in the plaquette and intertwiner labellings. Following the description of 
an explicitly local amplitude in the case of $D=3$ and $G=SU(2)$ by Anishetty \emph{et~al.} in~\cite{Anishetty90, 
Anishetty91, Anishetty93}, Halliday and Suranyi observed in~\cite{Halliday} that a local amplitude in four 
and higher dimensions can in principle be defined. However, a practical formula for the dual amplitude 
in $D=4$  (analogous to that of~\cite{Anishetty90, Anishetty91, Anishetty93}) has proven more elusive.

Recently, a systematic approach to non-abelian dual theories that emphasizes the underlying 
represen\-ta\-tion-theoretic structure has been developed starting with work by Oeckl and
Pfeiffer~\cite{OecklDGT, OecklPfeiffer}.  These developments make use of the diagrammatic
formalism to establish explicit representation-theoretic formulae for the local amplitudes 
that arise upon expanding in an intertwiner basis.   

It was demonstrated in~\cite{CCK} that numerical computations could be 
carried out for the dual form of LGT with $G=SU(2)$ on a cubic 
three-dimensional ($D=3$) lattice. 
One major hurdle to extending the algorithm of~\cite{CCK} to $D=4$ is
the problem of finding a closed-form formula for the vertex amplitude that can be evaluated
efficiently on a computer and gives reasonable scaling with increasing spin labels.

In what follows, we present such an explicit and efficient
evaluation of the vertex amplitude, a non-trivial function of 48 spin labels
that dominates the computational costs of evaluating the dual amplitude 
in four dimensions. Our implementation of this formula has been validated for a range of special cases;
currently, full Monte Carlo simulations using this amplitude are under way and will be described in
an upcoming paper.

\section{Explicit algorithm for the vertex amplitude}
\subsection{Review}
The steps we shall use to find an explicit formula for the vertex
amplitude are analogous to those described in~\cite{CCK} for the case
of $D=3$.  For a hypercubic lattice $\kappa$ in four dimensions,
we write the integral of the six $SU(2)$ matrices sharing a given edge variable $g_e$ as follows:
\begin{equation}\label{eq:HIdef}
  \int dg_e \, U_{j_1}(g_e)^{b_1}_{a_1} \, U_{j_2}(g_e)^{b_2}_{a_2} \,
               U_{j_3}(g_e)^{b_3}_{a_3} \, U_{j_4}(g_e^{-1})^{b_4}_{a_4} \, 
               U_{j_5}(g_e^{-1})^{b_5}_{a_5} \, U_{j_6}(g_e^{-1})^{b_6}_{a_6}
  = \int dg_e~\raisebox{-1.5cm}{\includegraphics[height=3.4cm]{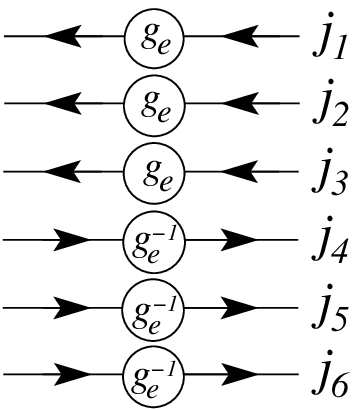}} \,\, ,
\end{equation}
where $j_1, \ldots, j_6$ denote irreducible representations of $SU(2)$, which we
index by non-negative half-integers $0, \frac{1}{2}, 1, \ldots$.
The RHS of~(\ref{eq:HIdef}) is a projection map from the space of all
linear maps $j_{6} \otimes j_{5} \otimes j_{4} \rightarrow j_{1} \otimes j_{2} \otimes j_{3}$
onto the space of intertwiners (maps commuting with the action of $G$), and can 
thus be expanded using an orthogonal basis of intertwiners $I_i: j_{6} \otimes j_{5}
\otimes j_{4} \rightarrow j_{1} \otimes j_{2} \otimes j_{3}$ as follows:
\begin{equation}\label{eq:HIresolution}
%	\raisebox{-0.81638cm}{\includegraphics[height=1.7328cm]{Hproj}}
\int dg_e~\raisebox{-1.45cm}{\includegraphics[height=3.4cm]{cables}} 
	~=~\sum_i \frac{I_i I^*_i}{\langle I^*_i,I_i\rangle}
	~=~\sum_i
		\frac{
			\raisebox{-0.169cm}{\includegraphics[height=2.60cm]{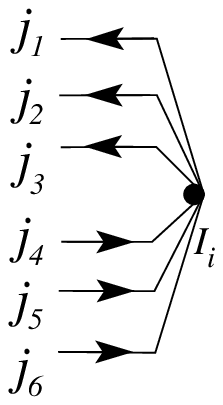}} ~~
			\includegraphics[height=2.4cm]{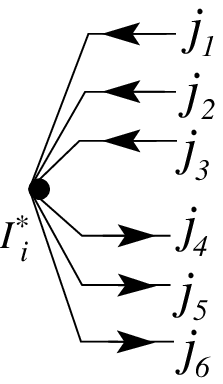}
		} {\raisebox{0cm}{\includegraphics[height=2.2cm]{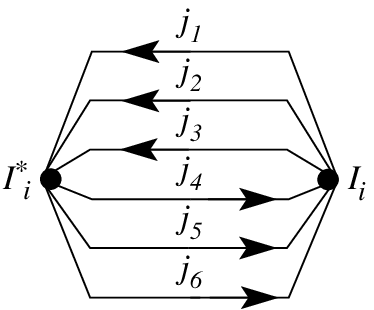}}}.
\end{equation}

\begin{figure}[htb]
% The trim command here enlarges the bounding box a bit, since it
% is too small in the .eps file.
\includegraphics[scale=0.54,trim=-2 0 -5 0]{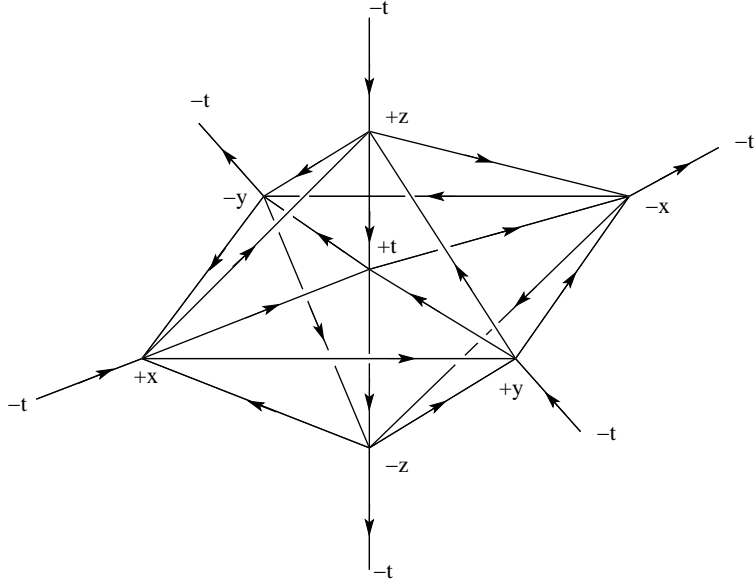}
\caption{Vertex amplitude for hypercubic lattice in four dimensions.}
\label{fig:48j}
\end{figure}

Carrying out the contractions results in a spin foam model (see, for 
example,~\cite{OecklPfeiffer,CCK,Conrady} for details), with the vertex 
amplitude defined as a spin network evaluation of the network shown in Figure~\ref{fig:48j}.
In drawing this figure, we have placed the vertex labelled $-t$ ``at infinity''. 
The edges of this network carry the irrep labels $j_i$ from the associated plaquettes meeting
at the vertex and the vertices carry the intertwiner labels appearing in the expansion~(\ref{eq:HIresolution}),
associated to the edges incident to the vertex.
The orientations of the edges are determined from an
orientation of the plaquettes in the lattice. 
We note in passing that
this network has 24 edges and 8 vertices, and shares the same graph structure as a
``hyperoctahedron''. 

To apply diagrammatic techniques leading to an explicit formula, our first task is to 
work with a diagram possessing 3-valent vertices, for which recoupling rules
from~\cite{CFandS, KandL} can be applied.  A basis for the 6-valent intertwiners appearing in
(\ref{eq:HIresolution}) can be readily chosen in terms of 
3-valent networks having six edges entering or leaving (see Figure~\ref{fig:splitting}). Because 
three internal edges are required, the edges  of the four-dimensional lattice will carry 
three half-integer spin labels, subject to some constraints. 
(In contrast, for the 4-valent vertices arising in $D=3$, only 
a single label is required to define the intertwiner assigned to an edge.)

Returning to the spin network that defines the vertex amplitude, we see that
because each network vertex is associated with three spins, the vertex amplitude
becomes a function of 48 spins in total; 24 from the network edges (coming from 
lattice plaquettes) and 24 from the 8 network vertices (3 from each of the 8 lattice edges).  Thus, 
we shall refer to a vertex amplitude (with its vertex splitting specified) as
a \emph{$48j$ symbol}.  It should be noted that different patterns of splittings over 
the edges of the lattice can be chosen leading to numerically different vertex 
amplitudes\footnote{Of course, for a given labelling of plaquettes, the sum over intertwiner labels
of the product of edge and vertex amplitudes over the lattice is independent of the choice of splitting.}.

An important technical consideration is that the network resulting
from an arbitrary splitting does not necessarily have translation symmetry over all the lattice vertices, as the splitting along one direction may not be related to the splitting in its opposite direction by a translation of the network. Indeed, this
is the case for the choice of splitting we will introduce below.
However, a pattern of partial reflections leading to $2^4=16$ ``versions'' of the $48j$ symbol can be
repeated to fill a lattice\footnote{The lattice must have even side-lengths, which is not a serious constraint in practice.}.   
As the sixteen versions are related to each other by appropriate permutations of arguments, it is sufficient to work out a
single case, which we shall by convention define to be the $48j$ symbol at the origin\footnote{A similar
phenomenon can arise in the three-dimensional case~\cite{Dass94},  although there it is sufficient to use 
two closely related network evaluations; alternatively, a translation invariant splitting can be used~\cite{CCK}.}.

In what follows we present the explicit evaluation of  the $48j$ symbol associated with a particular splitting 
(choice of intertwiner basis) that we give in  the next section. 

\subsection{Step 1 --- A splitting}
\begin{figure}[htb]
\includegraphics[scale=0.66]{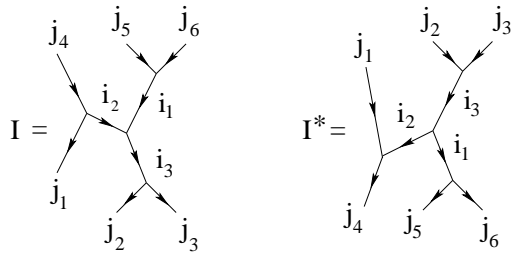}
\caption{Our choice of 6-valent intertwiners: a ``three-fold'' splitting.}
\label{fig:splitting}
\end{figure}

\begin{figure}[htb]
\includegraphics[scale=0.5]{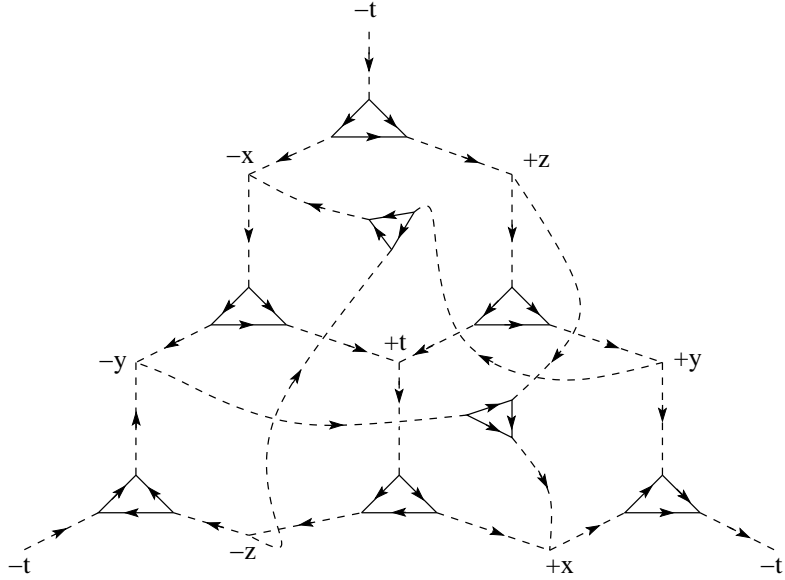}
\caption{A spin network diagram representing the $48j$ symbol in our choice
of splitting.}
\label{fig:48j_split}
\end{figure}

We now specify the splitting we shall use for our 6-valent intertwiners. 
First we recall that a 3-valent intertwiner is uniquely determined up to
a scale factor, and to fix conventions we use the intertwiners defined
in~\cite{CFandS}.
Figure~\ref{fig:splitting} shows a 6-valent intertwiner $I$ and its dual $I^{*}$.
As the internal labels $i_1$, $i_2$ and $i_3$ vary, the intertwiners $I$
form an orthogonal basis of all 6-valent intertwiners, and these are the splittings
used in our recoupling of the vertex amplitude below. Observe 
here that the six incident edges are grouped into three pairs; the choices 
we make are shown in Figure~\ref{fig:48j_split}, 
in which each 6-valent vertex has been replaced by an intertwiner $I$ as in
Figure~\ref{fig:splitting}.  The central vertex of $I$ is labelled with the
same label as the original 6-valent vertex had.

Before we can apply recoupling rules to explicitly compute the vertex
amplitude spin network, we must cast it into the form where the rules 
of~\cite{CFandS,KandL} can be applied.  To do this we follow the procedure described 
in Appendix A of~\cite{CCK}, which relates the original spin network (with oriented edges) defining 
the $48j$ symbol to a spin network in the sense of~\cite{CFandS,KandL};
although the two are equal in magnitude, in general there is a sign factor that 
depends on the spin arguments. 

\begin{figure}[htb]
\includegraphics[scale=0.40, trim=0 0 0 -5]{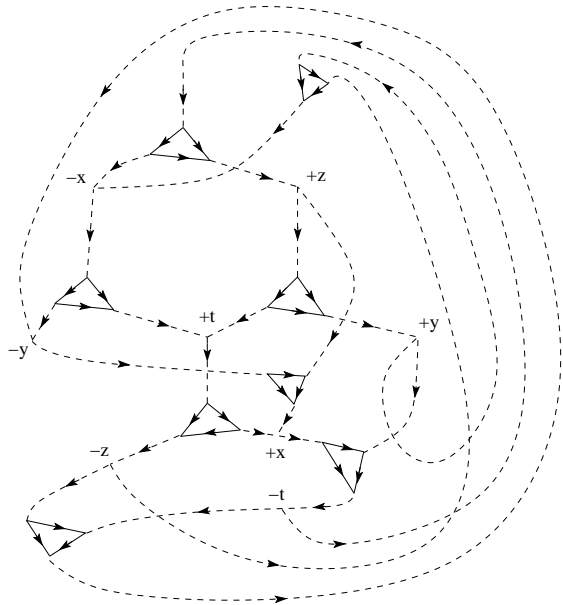}
\caption{Redrawing of vertex amplitude to determine sign factor.}
\label{fig:orderedmap}
\end{figure}

To determine the sign factor, we start by redrawing the network with all edges directed down the page, except
for some edges which wrap around the right to the top of the page and contribute a sign
factor.   In doing this, the order of the edges incident on each vertex must be kept the same,
but over-crossings and under-crossings can be reversed.
Such a mapping is presented in Figure~\ref{fig:orderedmap}.
We first recall each line wrapping around contributes a sign factor of $(-1)^{2 i}$. 
We next observe that each intertwiner label that appears in the overall sign factor $(-1)^{2 \sum i}$
for a given vertex will also appear in the sign factor of the vertex on the other side of the edge.  Thus,
globally these sign factors will cancel.  Because of this, we are free to ``drop the orientation'' of 
the original spin network and apply the rules of~\cite{CFandS} and~\cite{KandL}, as was done in~\cite{CCK}.

\subsection{Step 2 --- Collapse of triangles}
Referring to the splitting we have indicated in Figure~\ref{fig:48j_split}, we have  drawn the diagram
so as to make clear eight triangles, the edges of which are labelled by spins coming from plaquettes, and the vertices of which connect to edges carrying intertwiner labels.   The next step in our recoupling procedure is to collapse each of these by applying the identity 
\begin{eqnarray}
\begin{matrix}\includegraphics[scale=0.59]{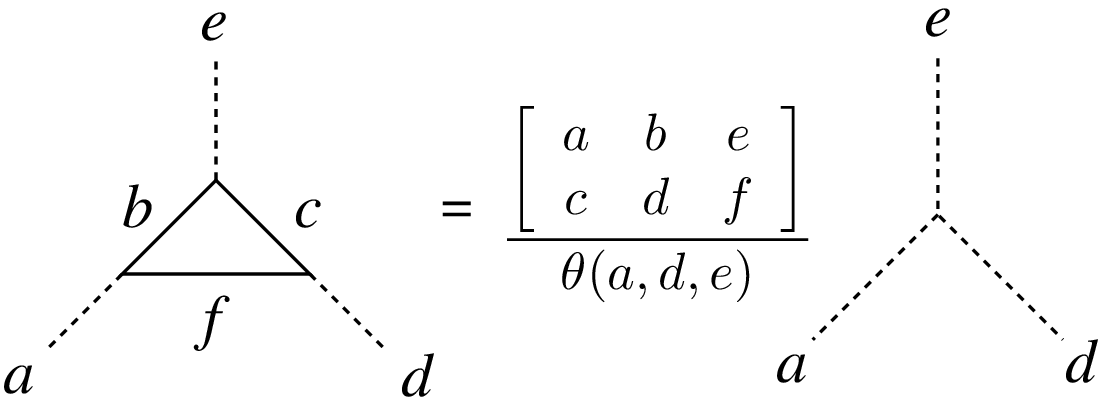}\end{matrix},
\label{eqn:collapse}
\end{eqnarray}
which introduces the Kauffman-Lins tet symbol
(the square bracketed expression) and theta network
\footnote{As discussed in~\cite{CCK}, the tet network is closely related to the Racah-Wigner $6j$ symbol.
Note that~\cite{KandL} uses integer spins where we and~\cite{CFandS} use half-integer spins.}%
.

\begin{figure}[htb]
\includegraphics[scale=0.47, trim=0 0 0 -3]{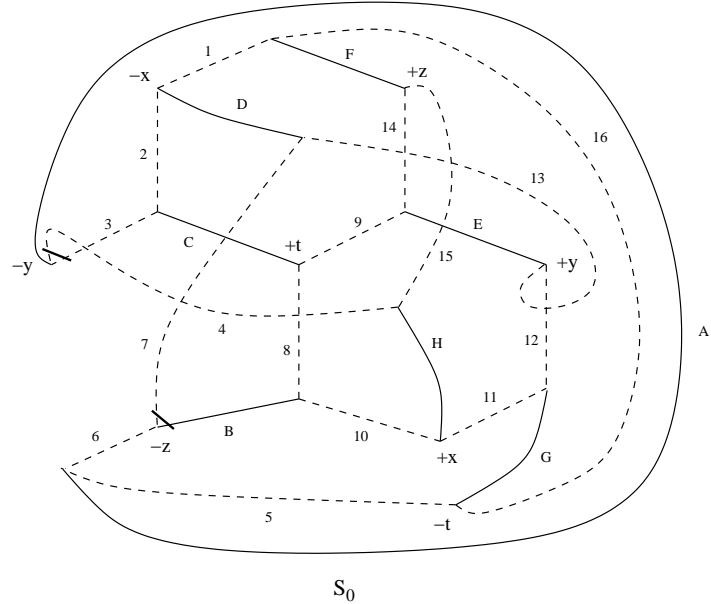}
\caption{Spin network evaluation after removing triangles. In Step 3, recoupling
moves are applied to solid lines.}
\label{fig:recoupling1}
\end{figure}

After collapsing the triangles, the vertex amplitude $A_{v}$ is given by
\begin{equation}
A_{v}[j_{1},j_{2},\ldots,j_{24},i_{1},i_{2},\ldots,i_{24}]= \tau_{1} \tau_{2}
\left(
\prod_{n=1}^{8}
\frac{
\left[ 
\begin{array}{ccc}
i_{a_{1}(n)}   &  j_{b_{1}(n)}  &  i_{e_{1}(n)}  \\ 
j_{c_{1}(n)}   &  i_{d_{1}(n)}  &  j_{f_{1}(n)}
\end{array} 
\right]}
{
\theta(i_{a_{1}(n)},i_{d_{1}(n)},i_{e_{1}(n)}) 
}
\right)S_{0} ,
\label{eq:step2result}
\end{equation}
where $S_{0}$ is the spin network depicted in Figure~\ref{fig:recoupling1}.  
In Figure~\ref{fig:recoupling1} it should be noted that, relative to Figure~\ref{fig:orderedmap}, two twists (indicated by bars spanning twisted strands) have been introduced, giving rise to associated sign factors $\tau_{1}$ and $\tau_{2}$ (see Step 3 below for discussion).    This was to put the diagram in a particular planar order so that
the identity (\ref{eqn:recouple}) defined in the next section below can be applied unambiguously. 

An interesting remark at this point is that the dependence on the 24 plaquette spins $j_i$ has
 been completely absorbed into the product of collapse factors in~(\ref{eq:step2result});
the remaining spin network is purely a function of the 24 intertwiner labels.

\subsection{Step 3 --- Transforming to a ladder}

\begin{figure}[htb]
\includegraphics[scale=0.4,trim=0 0 -3 0]{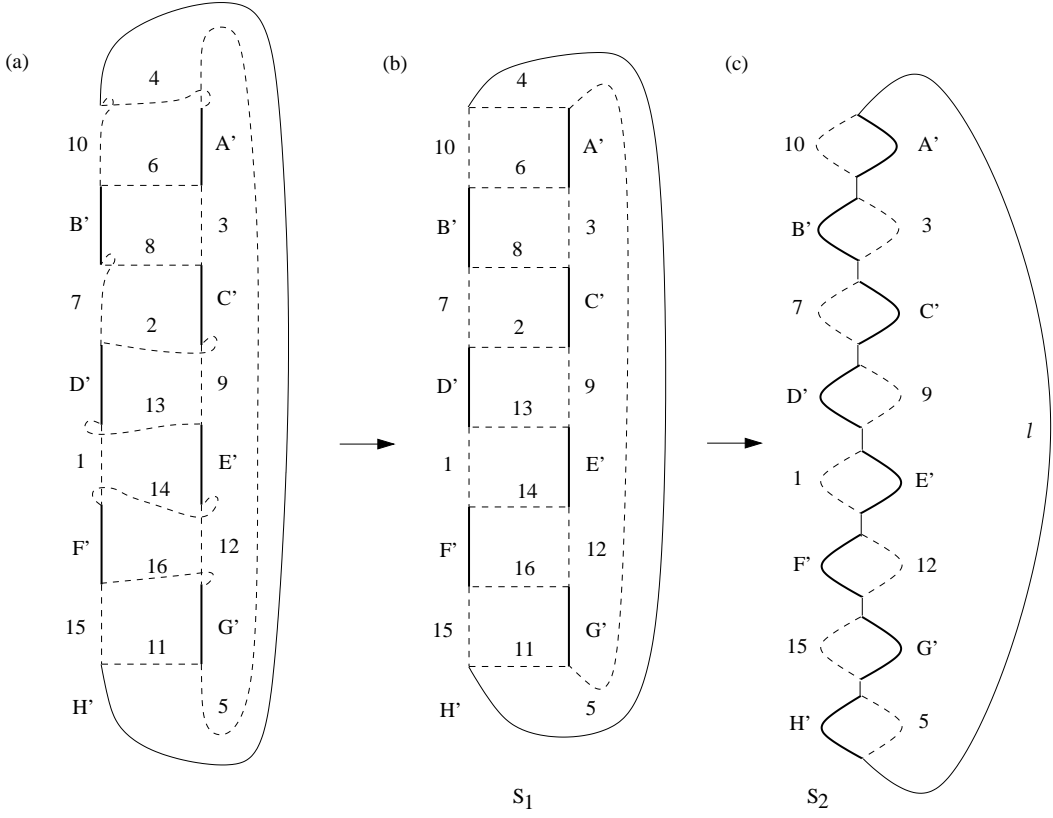}
\caption{Final steps in recoupling, leading to a matrix multiplication of 
tet networks.}
\label{fig:recoupling_final}
\end{figure}

In this step we transform the spin network $S_{0}$ into a ladder via  a sequence of eight recoupling moves 
of the form
\begin{eqnarray}
\begin{matrix}\includegraphics[scale=0.4]{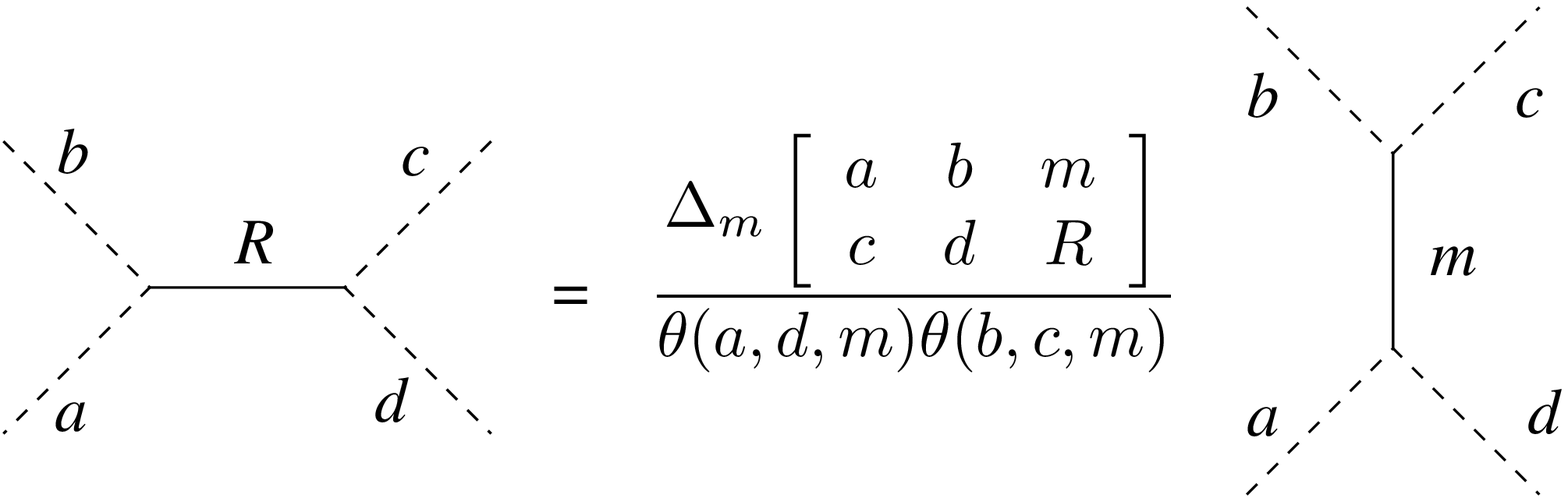}\end{matrix},
\label{eqn:recouple}
\end{eqnarray}
where $\Delta_m = (-1)^{2m}(2m+1)$.
The moves are indicated in Figure~\ref{fig:recoupling1} as follows.  
The eight edges to be recoupled are drawn as solid lines and labelled with
letters from $A$ through $H$.  
As given by~(\ref{eqn:recouple}), each of these moves leads to an auxiliary sum over an additional irrep label $m_i$. 
Upon applying the recoupling moves one arrives at the spin network shown 
in Figure~\ref{fig:recoupling_final}(a).

While these moves are sufficient to give a network having the topology of a ladder, there are a number of vertices
at which ``twist moves'' have to be performed to arrive at a ladder without crossings.  These twists do not effect
the magnitude of the spin network evaluation but can introduce signs, as follows:
\begin{eqnarray}
\begin{matrix}\includegraphics[scale=0.52]{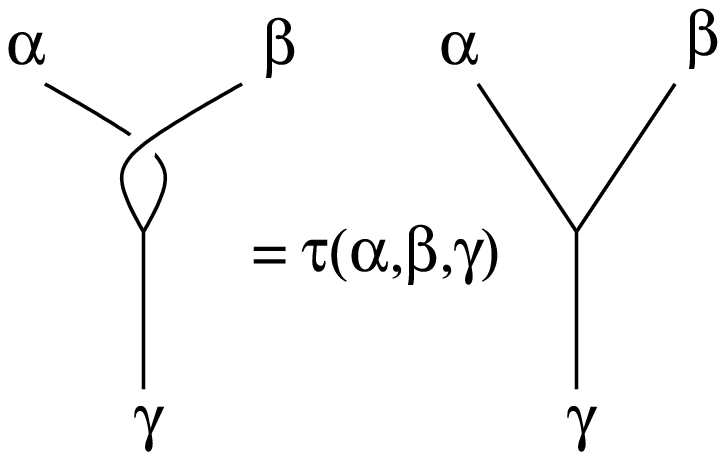}\end{matrix},
\label{eqn:twistmove}
\end{eqnarray}
where $\tau(\alpha,\beta,\gamma)=(-1)^{\alpha+\beta-\gamma}$.
The resulting expression for $S_{0}$ is
\begin{equation}
S_{0}[i_{1},i_{2},\ldots,i_{24}]= 
\sum_{m_{1}}\sum_{m_{2}}\cdots\sum_{m_{8}}
\left(
\prod_{n=1}^{8}
\frac{
\Delta_{m_{n}}
\tau_{n}
\left[ 
\begin{array}{ccc}
i_{a_{2}(n)}   &  i_{b_{2}(n)}  &  m_{n}  \\ 
i_{c_{2}(n)}   &  i_{d_{2}(n)}  &  i_{R(n)}
\end{array} 
\right]}
{\theta(i_{a_{2}(n)},i_{d_{2}(n)},m_{n})\theta(i_{b_{2}(n)},i_{c_{2}(n)},m_{n}) }
\right)S_{1} ,
\label{eqn:step2recouple}
\end{equation}
where $\tau_{n}=\tau (i_{\alpha(n)}, i_{\beta(n)}, m_n)$ are the factors associated with the twists that are attached to the eight recoupled edges. $S_{1}$  is the ladder spin network shown in Figure~\ref{fig:recoupling_final}(b).
The spins $m_1$ through $m_8$ are associated to the new edges, labelled $A'$ through $H'$.
The eight intertwiners that labelled the edges $A$ through $H$ in Figure~\ref{fig:recoupling1},
denoted $R$ in the recoupling move~(\ref{eqn:recouple}), appear as $i_{R(n)}$
in~(\ref{eqn:step2recouple}).

\subsection{Step 4 --- Collapsing the ladder}
Here we apply the recoupling move to the eight ``rungs'' of the ladder.  Superficially, this leads to an additional eight
auxiliary summations. However, upon removing the bubbles (which are proportional to the identity intertwiner) it becomes clear that all eight auxiliary spins must be equal for the resulting network to be non-zero (Schur's Lemma). Thus, only a \emph{single} additional auxiliary summation is required.  Hence
\begin{equation}
S_{1}[i_{1},i_{2},\ldots,i_{16},m_{1}, \ldots, m_{8}]=
\sum_{l}
\left(
\prod_{n=1}^{8}
\frac{\Delta_{l}
\left[ 
\begin{array}{ccc}
i_{a_{3}(n)}   &  m_{b_{3}(n)}  &   l \\ 
i_{c_{3}(n)}   &  m_{d_{3}(n)}  &  i_{r(n)}
\end{array} 
\right]}
{\theta(i_{a_{3}(n)},m_{d_{3}(n)},l)\theta(i_{b_{3}(n)},m_{c_{3}(n)},l) }
\right)S_{2} ,
\label{eqn:laddercollapse}
\end{equation}
where $S_2$ is the chain of bubbles; note here that the $i_{r(n)}$ are the intertwiners corresponding
to the ``rungs'' of the ladder in Figure~\ref{fig:recoupling_final}(b).  

To complete the evaluation we apply the ``bubble collapse'' move 
\begin{eqnarray}
\begin{matrix}\includegraphics[scale=0.56]{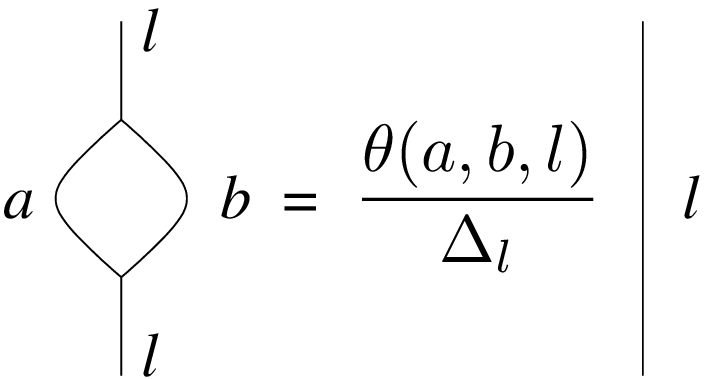}\end{matrix}
\label{eqn:bubble}
\end{eqnarray}
to each of the eight bubbles in Figure~\ref{fig:recoupling_final}(c),
which results in a product of theta and delta factors:
\begin{equation}
S_2[i_{1},i_{2},\ldots, i_{8}, m_{1}, \ldots, m_{8}] = \Delta_{l}\left( \prod_{n=1}^{8} \frac{ \theta(i_{n},m_{n},l) } {\Delta_{l}}\right) .
\label{eqn:bubblecollapse}
\end{equation}
The $\Delta_l$ factor in front of the product comes from the loop formed after transforming the eight bubbles.

Assembling the results of our previous steps yields the following:
\begin{multline}
A_{v}[j_{1},j_{2},\ldots,j_{24},i_{1},i_{2},\ldots,i_{24}]=
\left(
\prod_{n=1}^{8}
\left[ 
\begin{array}{ccc}
i_{a_{1}(n)}   &  j_{b_{1}(n)}  &  i_{e_{1}(n)}  \\ 
j_{c_{1}(n)}   &  i_{d_{1}(n)}  &  j_{f_{1}(n)}
\end{array} 
\right] 
\right)
\sum_{l}
\sum_{m_{1}}\sum_{m_{2}}\cdots\sum_{m_{8}}
\\ 
\Delta_{l}
\Bigg(
\prod_{n=1}^{8}
\frac{
\Delta_{m_{n}}\tau_{n}
\left[ 
\begin{array}{ccc}
i_{a_{2}(n)}   &  i_{b_{2}(n)}  &  m_{n}  \\ 
i_{c_{2}(n)}   &  i_{d_{2}(n)}  &  i_{R(n)}
\end{array} 
\right]}
{\theta(i_{a_{2}(n)},i_{d_{2}(n)},m_{n})\theta(i_{b_{2}(n)},i_{c_{2}(n)},m_{n}) }
\frac{
\left[ 
\begin{array}{ccc}
i_{a_{3}(n)}   &  m_{b_{3}(n)}  &   l \\ 
i_{c_{3}(n)}   &  m_{d_{3}(n)}  &  i_{r(n)}
\end{array} 
\right]}
{\theta(i_{a_{3}(n)},m_{d_{3}(n)},l)\theta(i_{b_{3}(n)},m_{c_{3}(n)},l) }
 \theta(i_{n},m_{n},l) 
 \Bigg),
\label{eqn:FinalFormula}
\end{multline}
where the $\Delta_l$ factors appearing in the denominator of~(\ref{eqn:bubblecollapse}) cancel with the $\Delta_l$ factors in~(\ref{eqn:laddercollapse}).

\subsection{Computational Cost}

Suppose we choose a spin cut-off of $j$, so all spin and intertwiner
labels must be at most $j$.  We will estimate the time required to
compute the vertex amplitude~(\ref{eqn:FinalFormula}).

The factors in~(\ref{eqn:FinalFormula}) each depend on only one $m_{n}$ variable,
except for the third tet network, which depends on two of the $m_n$ variables.
In fact, one can show that this dependence is in a cyclical order, 
so that, for fixed $l$, the vertex amplitude can be computed as the trace of a 
product of eight matrices.
The range of the $m_{n}$ variables is restricted by the triangle equalities to be $O(j)$,
so the matrices have size $O(j) \times O(j)$.  The entries of these matrices are
products of tet networks and $\theta$ networks, which require $O(j)$ time to compute.
Thus forming the matrices and multiplying them together each require $O(j^3)$ time.
Considering the sum over $l$, which is also constrained to be $O(j)$, we
conclude that the overall cost to computing the vertex amplitude is $O(j^4)$. 

For comparison with the $D=3$ case, we recall that the amplitude introduced
by~\cite{Anishetty90,Anishetty91,Anishetty93} and used in~\cite{Dass94,Dass83,DassShin} 
was  $O(j)$. This $O(j)$ algorithm as well as an alternative $O(j^2)$ vertex amplitude
were used in the $D=3$ dual LGT simulations~\cite{CCK}.

\subsection{Edge amplitude}
\begin{figure}[htb]
\includegraphics[scale=0.58]{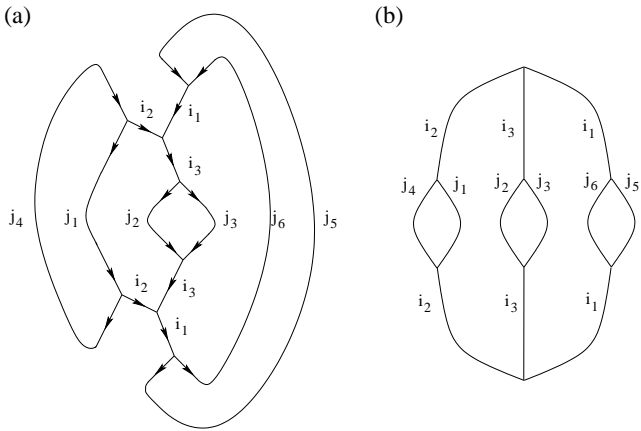}
\caption{Normalization of edges.}
\label{fig:edgenorm}
\end{figure}

To complete the model, we describe how the normalization factor appearing in 
the denominator on the RHS of~(\ref{eq:HIresolution}) can be given an explicit expression. 
Having chosen the three-fold splitting described above for our intertwiners, we  
draw the edge amplitude evaluation as shown in Figure~\ref{fig:edgenorm}(a).   
As above, one can compute the sign relating the original network with directed edges 
to an undirected Kauffman-Lins spin network, and once again one finds that globally
these signs cancel.  The undirected spin network is shown 
in Figure~\ref{fig:edgenorm}(b) and can be transformed into a theta network by 
collapsing the three bubbles according to the identity~(\ref{eqn:bubble}), as follows:
\begin{align}
&A_{e}[j_{1},j_{2},j_{3},j_{4},j_{5},j_{6},i_{1},i_{2},i_{3}]=
\frac{\theta(j_4,j_1,i_2) \theta(j_2,j_3,i_3) \theta(j_6,j_5,i_1)} {\Delta_{i_1}\Delta_{i_2}\Delta_{i_3}} \theta(i_1,i_2,i_3).
\end{align}
Here $j_1, \ldots, j_6$ are the labels for the six plaquettes incident to an edge and $i_1, i_2, i_3$ are the 
three intertwiner labels.
Figure~\ref{fig:edgenorm} also makes it clear that different intertwiners are orthogonal.

\subsection{Validation of the Vertex and Edge Amplitudes}
The best comprehensive check for the validity
of the amplitude derived above is a full simulation of the dual theory at increasingly
large cut-offs. While this work is currently in progress, a more limited
form of validation has been done.  In the case where dual configurations are ensembles of closed, self-avoiding
surfaces on the lattice, there is an explicit closed-form expression for the resulting amplitude related 
to the coloring of self-avoiding surfaces by irreps and their Euler characteristic; this formula is derived 
in~\cite{Conrady}. To verify that the amplitude was correct for these configurations, code was implemented to generate a range 
of such self-avoiding topologies (corresponding topologically to 2-tori and 2-spheres) and colorings. 
Using an implementation of the vertex and edge amplitudes given above, the amplitudes of self-avoiding dual configurations
were found to be in agreement with the closed form formula. 

\section{Conclusions}
We have presented an explicit formulation of pure Yang-Mills theory on a $D=4$ hypercubic lattice
with $G=SU(2)$. Computationally, the critical part of the amplitude is the vertex amplitude, a function of 
48 spins coming from the 24 plaquette labels and 24 intertwiner labels associated to a vertex. 
In a variety of special cases where a closed form formula was available, the amplitude was successfully verified.  
The first results of dual simulations using this amplitude will be reported on in forthcoming work by the first 
author.

While the vertex amplitude scales with $j^4$ rather than $j$ (as in the $D=3$ case), the achievable performance is sufficient 
to enable the first practical four-dimensional computations within the dual model.  This is an important step in evaluating
the feasibility of dual algorithms for non-abelian LGT as a possible alternative to conventional methods. 
We would also like to remark that the discovery of the original $D=3$ amplitude~\cite{Anishetty90,Anishetty91,Anishetty93}
prompted a number of intriguing analytic results.  These include work by Diakonov and Petrov~\cite{DiaPet99} revealing a 
gravitational interpretation to the dual degrees of freedom, as well as recent work by Conrady~\cite{ConradyGluons}.   
With the present result,  the way is now clear to establish analogous analytic results in $D=4$.  In particular, it will 
be interesting to see which results of the above work are specific to $D=3$ and which ideas can be extended $D=4$.
 
\section{Acknowledgements}
The authors would like to thank Florian Conrady and Igor Khavkine for useful discussion relating to this work. 
Both authors are funded by NSERC. This work was made possible by the facilities of the Shared Hierarchical 
Academic Research Computing Network (SHARCNET).

\end{document}